\documentstyle[12pt]{article}

\begin{document}

\title{Covariant Hamiltonian formalisms for particles and antiparticles}
\author{Edgardo T. Garcia Alvarez$^1$ and Fabi\'an H. Gaioli$^1$}
\maketitle

\begin{abstract}
The hyperplane and proper time formalisms are discussed mainly for the
spin-half particles in the quantum case. A connection between these
covariant Hamiltonian formalisms is established. It is showed that choosing
the space-like hyperplanes instantaneously orthogonal to the direction of
motion of the particle the proper time formalism is retrieved on the mass
shell. As a consequence, the relation between the St\"uckelberg-Feynman
picture and the standard canonical picture of quantum field theory is
clarified.

\vspace{1.0in}

$^1$Instituto de Astronom\'\i a y F\'\i sica del Espacio, C.C. 67, Suc. 28,
1428 Buenos Aires, Argentina and Departamento de F\'\i sica, Facultad de
Ciencias Exactas y Naturales, Universidad de Buenos Aires, 1428 Buenos
Aires, Argentina. E-mail: gaioli@iafe.uba.ar.
\end{abstract}

\section{Introduction}

The unification of the principles of relativity and quantum mechanics
presents a serious obstacle. On the one hand, as from the seminal work of
Minkowski (1908) the first theory deals with the space and time on an equal
footing:

\smallskip\ 

``Space by itself, and time by itself, are doomed to fade away into mere
shadows, and only a kind of union of the two will preserve an independent
reality.''

\smallskip\ 

On the other hand, the principles of quantum mechanics, originally developed
in a canonical formalism, have broken this symmetry by choosing the
coordinate time of a given frame of reference to label the evolution of the
system. Therefore the standard canonical formalism does not provide a
relativistic invariant description of the dynamical evolution of the system.
Moreover in this framework it is not possible to describe simultaneously
particles and antiparticles at a first quantized level.

Different covariant formalisms were proposed to overcome these obstacles.
Significative advances were obtained when the problem was faced with the
purpose of reformulating the old fashioned canonical theory of quantum
electrodynamics. Two kinds of solutions were presented at the Pocono
conference half a century ago (Schweber, 1986; Mehra, 1994). One by
Schwinger (1948), who, as Tomonaga (1946), essentially generalized the
standard canonical formalism of quantum field theory to arbitrary space-like
surfaces, and another one, containing more radical changes, by Feynman
(1951) (see also Schweber, 1986 and Mehra, 1994). Feynman's ideas, like
St\"uckelberg's (1941a, b, 1942) ones, dealt with the space-time
trajectories of charged particles, and were essentially formulated at a
first quantized level.

In this work we recall such ideas in order to discuss the covariant
Hamiltonian formalism for relativistic particles. We shall focus our
exposition on the Fleming hyperplane formalism (Fleming, 1965, 1966),
closely related to the Tomonaga-Schwinger ideas,\footnote{%
See Jauch and Rohrlich (1976) for a list of references about formalisms
involving arbitrary space-like surfaces.} and the Feynman proper time
formalism, with the aim of establishing a bridge between them. This will
clarify many misunderstood issues of the connection between the standard
canonical picture and the Feynman space-time picture, from the Feynman point
of view.\footnote{%
From the canonical point of view, this connection was established by Dyson
(1948).}

The main purpose of this work is to discuss Feynman's formalism for the
relativistic spin-$\frac 12$ particles in the quantum case (Sections 4\ and
5), but for pedagogical reasons we begin discussing the problem at the
classical level for the spinless relativistic particle in Sections 2 and 3.

Throughout this work we use natural units ($\hbar =c=1).$ Our convention for
the metric is

\begin{equation}
ds^2=\eta _{\mu \nu }dx^\mu dx^\nu ,\ \hspace{0.3in}\eta _{\mu \nu }=%
{\rm diag}(1,-1,-1,-1).  \label{ds2}
\end{equation}

\section{The classical relativistic particle in the hyperplane}

The standard canonical formalism usually considers particle states only. Let
us go beyond this formalism by using the Hamiltonian

\begin{equation}
H=\epsilon \sqrt{m^2+{\bf p}^2},  \label{H}
\end{equation}
where $\epsilon $ is $+1$and $-1$ for particles and antiparticles
respectively. In this way we adopt St\"uckelberg's (1941a, b, 1942) and
Feynman's (1948, 1949) ideas introducing the concept of antiparticles at the
classical level as negative energy states going backwards in coordinate time.%
\footnote{%
Such a notion can be also used for deriving the Dirac equation from first
principles (Gaioli and Garcia Alvarez, 1995).} As we have stressed above,
the canonical formalism privileges the temporal coordinate $x^0$ of a
reference frame in order to describe the evolution, the one in which $H$ is
the temporal component of the four-momenta $p_\mu =(H,-{\bf p)}$. In other
words, for each reference frame we have a different Hamiltonian which
generates the dynamical evolution of the system in the corresponding
coordinate time. The key idea of the hyperplane formalism is unifying such
multiplicity of dynamical descriptions by taking the temporal coordinate of
a privileged frame, $\tau ,$ for labeling the dynamics. This choice can be
written in an invariant language as follows: Space-time is foliated with a
family of space-like surfaces $n^\mu x_\mu -\tau =0$ (the hyperplanes)
characterized by a normalized vector $n^\mu $ ($n^\mu n_\mu =1)$ orthogonal
to the surfaces

\begin{equation}
n^\mu =\frac{\partial x^\mu }{\partial \tau }.  \label{ndx}
\end{equation}
We have chosen the normal vector $n^\mu $ in the direction of the future
light cone. Following this convention the components of the temporal vector $%
n^\mu $ adopt the simple form $(n_\tau )^\mu =(1,0,0,0)$ in the coordinates $%
(x_\tau )^\mu =(\tau ,{\bf x}_\tau )$ of the privileged frame$.$ Of course,
the choice of such privileged foliation, in order to label the dynamics, is
arbitrary. Usually, each observer adopts his canonical foliation, that is
the one corresponding to coordinate time. But at this point the hyperplane
formalism dissociates the dual (geometrical and dynamical) role of the
temporal coordinates of the different reference systems. Each temporal
coordinate retains its geometrical role, but only one (arbitrarily chosen)
adopts the dynamical one. Note that in this sense the time $\tau $
registered by the privileged coordinate frame is an absolute scalar
parameter. That is, any lapse of $\tau (P_1)-\tau (P_2)$ corresponding to
two events occurred at points $P_1$ and $P_2$ in space-time is (by
definition) independent of the coordinate systems chosen.\footnote{%
What is relative is the lapse in the time cooordinates, i.e. for two systems 
$S$ and $S^{\prime },$ $x^0(P_1)-x^0(P_2)\neq $ $x^{0\prime
}(P_1)-x^{0\prime }(P_2)$.} As a consequence its conjugate variable, the
Hamiltonian $H(n),$ is also scalar, which becomes evident writing it as%
\footnote{%
Notice also that particle and antiparticle concepts are interchanged if we
reverse the direction of $n^\mu .$}

\begin{equation}
H(n)=n^\mu p_\mu .  \label{Hn}
\end{equation}

In this way the hyperplane formalism describes the evolution of the system
from any coordinate system in a covariant way. However, note that the
multiplicity of dynamical descriptions of the standard canonical formalism
discussed above was not lost. It is hidden in the arbitrary choice of $n^\mu
.$ The only thing that has been improved is that now the canonical formalism
is independent of the coordinate system chosen. That is the canonical
formalism provides a relativistic invariant description of the dynamical
evolution of the system.

Although the expression (\ref{Hn}) looks explicitly covariant, the canonical
formalism is rather complicated because of the variables $p_\mu $ are not
independent, since they satisfy the mass-shell constraint

\begin{equation}
p^\mu p_\mu =m^2.  \label{pp}
\end{equation}
An alternative expression for (\ref{Hn}) can be obtained in terms of the
canonical momentum ${\bf p}_\tau $ conjugated to the hyperplanes coordinates 
${\bf x}_\tau .$ Using (\ref{H}) in coordinates ($\tau ,{\bf x}_\tau )$ we
have 
\begin{equation}
H(n)=\epsilon \sqrt{m^2+p^\mu (n)p_\mu (n)},  \label{Hnpn}
\end{equation}
where\footnote{%
The expression (\ref{Hnpn}) was generalized to a curved space-time by Ferraro%
{\it \ et al.} (1987){\it .}}

\begin{equation}
p^\mu (n)=p^\mu -n^\mu (n^\nu p_\nu ),  \label{pn}
\end{equation}
is the four vector associated with ${\bf p}_\tau $ [which in the coordinates
of the privileged frame reads $(p_\tau )^\mu (n)=(0,{\bf p}_\tau )].$
However the new momentum variables do not simplify the problem, because they
also satisfy a constraint

\begin{equation}
p^\mu (n)n_\mu =0,  \label{cpn}
\end{equation}
since $p^\mu (n)$ is the projection of $p^\mu $ to the hyperplane $\tau =0$.

Notice also that the covariant Poisson brackets, $\left\{ f ,
g\right\} _{xp}\equiv \frac{\partial f}{\partial x^\alpha }\frac{\partial g}{%
\partial p_\alpha }-\frac{\partial g}{\partial x^\alpha }\frac{\partial f}{%
\partial p_\alpha },$ of (\ref{pn}) with the four-vector

\begin{equation}
x^\mu (n)=x^\mu -n^\mu (n^\nu x_\nu )  \label{xn}
\end{equation}
associated with ${\bf x}_\tau ,$ is not canonical

\begin{equation}
\left\{ x^\mu (n) , p^\nu (n)\right\} _{xp}=\eta ^{\mu \nu }-n^\mu
n^\nu .  \label{xnpn}
\end{equation}

We will return to this kind of problems later on.

Up to this point, we have not removed the undesirable arbitrariness in the
choice of $n^\mu .$ In the case of the free particle the only four-vector
that gives a privileged direction in the space-time is the four-velocity
(which for a spinless particle also coincides with the direction of its
four-momentum\footnote{%
This is not the case in general when the particle has spin. See, for
example, Corben (1961, 1968).}). We can remove the arbitrariness by choosing

\begin{equation}
n^\mu =\epsilon \frac{dx^\mu }{ds}=\epsilon \frac{p^\mu }m,  \label{ns}
\end{equation}
which identifies the canonical variable $\tau $ with the proper time of the
particle, 
\begin{equation}
\tau =\epsilon s,\hspace{0.3in}ds=\epsilon \sqrt{1-{\bf v}^2}dx^0,
\label{dsep}
\end{equation}
since this choice imposes our privileged system to be a system in which the
particle is at rest.\footnote{%
Note that $s$ is the proper time for both particles and antiparticles,
since, according to Stueckelberg, for antiparticles $dx^0<0.$} Using (\ref
{ns}) and the constraint (\ref{pp}) in (\ref{Hn}) we have

\begin{equation}
H(n)=\epsilon m=\epsilon \sqrt{p^\mu p_\mu },  \label{Hns}
\end{equation}
which shows that for particles the conjugate variable of $\tau $ becomes the
rest mass $m$.

\section{The proper time formalism for classical spinless particles}

Recently Hall and Anderson (1995) have proposed a covariant Hamiltonian
formalism for a relativistic particle based on a square root
super-Hamiltonian 
\begin{equation}
{\cal H}=\sqrt{p^\mu p_\mu },  \label{Hmoses}
\end{equation}
which resembles expression (\ref{Hns}) but with the four momentum $p^\mu $
not restricted to the mass shell (${\cal H}$ behaves as a positive variable
mass). Such a formalism originally developed by Moses (1969) and Johnson
(1969) and more recently discussed by Evans (1990), Hannibal (1991a), and us
(Aparicio {\it et al. }1995a, b) is a formalism free from constraints in
which the invariant evolution parameter is identified with the proper time.
In this framework, in contrast with (\ref{xnpn}), we have covariant
commutation relations 
\begin{equation}
\left\{ x^\mu , p^\nu \right\} _{xp}=\eta ^{\mu \nu }.  \label{xp}
\end{equation}
However, we must deal with an indefinite mass system,\footnote{%
Notice that (\ref{xp}) is incompatible with the mass-shell constraint (\ref
{pp}).} in such a way that the standard notion of definite-mass particles
and antiparticles are recovered specifying the initial conditions.

Hall's and Anderson's approach is interesting because, in spite of
postulating the form of the Hamiltonian, they derive ${\cal H}$ in a
constructive way by imposing physical requirements.

Their argument flows as follows: Let us assume a four-dimensional
Hamiltonian formalism whose equations of motion read

\begin{equation}
\frac{dx^\mu }{d\lambda }=\frac{\partial {\cal H}}{\partial p_\mu },%
\hspace{0.3in}\ \frac{dp^\mu }{d\lambda }=-\frac{\partial {\cal H}}{%
\partial x_\mu },  \label{eh}
\end{equation}
where $\lambda $ is an invariant evolution parameter and ${\cal H}={\cal H}%
(x,p)$ is the covariant Hamiltonian. Afterwards let us assume that the
invariant evolution parameter can be identified with the proper time,

\begin{equation}
\lambda =s.  \label{ls}
\end{equation}
Such a condition, in principle, allows us univocally to determine the form
of the free spinless Hamiltonian. In this case space-time homogeneity
imposes that ${\cal H}$ cannot explicitly depend on $x^\mu $, so ${\cal H}=%
{\cal H}(p^\mu )$, and the condition of being a scalar under Lorentz
transformations leads ${\cal H}$ to be an arbitrary function of $p=\sqrt{%
\eta _{\mu \nu }p^\mu p^\nu },$ the only scalar that we can form with the
available tensors of the theory. Using the equations of motion, the
constraint

\begin{equation}
\eta _{\mu \nu }\frac{dx^\mu }{ds}\frac{dx^\nu }{ds}=1,  \label{constr}
\end{equation}
can be rewritten as

\begin{equation}
\eta _{\mu \nu }\frac{\partial {\cal H}}{\partial p_\mu }\frac{\partial 
{\cal H}}{\partial p_\mu }=1,
\end{equation}
or, taking into account that $\frac{\partial {\cal H}}{\partial p_\mu }=%
\frac{d{\cal H}}{dp}\frac{p^\mu }p,$ as

\begin{equation}
\left( \frac{d{\cal H}}{dp}\right) ^2=1.  \label{deq}
\end{equation}
Finally, the differential equation (\ref{deq}) can be easily integrated to
give

\begin{equation}
{\cal H}=\pm p,  \label{J}
\end{equation}
which is the Moses-Johnson Hamiltonian. The four-velocity associated with
this Hamiltonian is

\begin{equation}
\frac{dx^\mu }{ds}=\pm \frac{p^\mu }p,  \label{dxJ}
\end{equation}
an equation which shows that for positive mass states we have $\lambda =s$,
with the sign specified in equation (\ref{dsep}).

Hall and Anderson have also generalized this argument for the case in which
the theory admits another four-vector $t^\mu ,$ giving in this case a
Hamiltonian of the type\footnote{%
Note that the new Hamiltonian is a particular case of the Hamiltonian (\ref
{Ht}) for $t^\mu =\pm p^\mu /p.$}

\begin{equation}
{\cal H}=t^\mu p_\mu .  \label{Ht}
\end{equation}
This ${\cal H}$ is admissible provided that the norm of $t^\mu $ satisfies

\begin{equation}
t^\mu t_\mu =1.  \label{tt}
\end{equation}

The equation of motion of the four-velocity results

\begin{equation}
\frac{dx^\mu }{d\lambda }=t^\mu ,  \label{dxst}
\end{equation}
so the constraint (\ref{constr}) is equivalent to condition (\ref{tt}). We
also remark the analogy of Hamiltonians (\ref{Hn}) and (\ref{Ht}). In the
conclusions of their work Hall and Anderson speculate with the possibility
of incorporating spin from such a generalization. At the end of this work we
show that this conjecture can be crystalized relaxing the normalization
condition (\ref{tt}), by choosing $t^\mu $ as the Dirac matrix $\gamma ^\mu
. $

\section{The Dirac equation in the hyperplane formalism}

In this Section we review the hyperplane formalism for a quantum spinning
particle described by the Dirac equation,

\begin{equation}
\gamma ^\nu i\partial _\nu \psi (x)=m\psi (x)  \label{D}
\end{equation}
(Hammer {\it et al., } 1968), with the aim of establishing a connection with
the proper time formalism in an analogous way to the one discussed at the
end of Section 2. Then our purpose is to translate the Hamiltonian form of
equation (\ref{D}),

\begin{equation}
i\partial _0\psi (x)=({\bf \alpha }.{\bf p}+\beta m)\psi (x)  \label{HD}
\end{equation}
$(p_\mu =i\partial _\mu ),$ into an arbitrary hyperplane.

There are two ways for doing it depending on whether we take equation (\ref
{D}) or equation (\ref{HD}) as a starting point. We begin discussing the
first possibility, proposed by Czachor (1995), which is more
straightforward. Multiplying both members of equation (\ref{D}) by $\gamma
^\nu $ and taking into account the identity $\gamma ^\mu \gamma ^\nu =\eta
^{\mu \nu }-i\sigma ^{\mu \nu },$ we have (K\'alnay and MacCotrina, 1968)

\begin{equation}
i\partial ^\mu \psi (x)=(i\sigma ^{\mu \nu }p_\nu +m\gamma ^\mu )\psi (x).
\label{KC}
\end{equation}
Now contracting equation (\ref{KC}) with $n^\nu ,$ we finally obtain
(Czachor, 1995)

\begin{equation}
i\frac{\partial \psi (x)}{\partial \tau }=n_\mu (i\sigma ^{\mu \nu }p_\nu
+m\gamma ^\mu )\psi (x),  \label{Cza}
\end{equation}
where we have used the chain rule and (\ref{ndx})

\[
\frac{\partial \psi }{\partial \tau }=\frac{\partial \psi }{\partial x^\mu }%
\frac{\partial x^\mu }{\partial \tau }=n^\mu \partial _\mu \psi . 
\]
The original derivation of Hammer {\it et al}. (1968) follows a similar
argument to the one used for obtaining expression (\ref{Hnpn}). It departs
from equation (\ref{HD}) in the privileged reference system and rewrites
this equation in a covariant way. Adapted to our notation and conventions,
the Hammer-MacDonald-Pursey equation reads:

\begin{equation}
\begin{array}{c}
i\frac{\partial \psi }{\partial \tau }=H(n)\psi , \\ 
\\ 
H(n)=\left[ \alpha ^\mu (n)p_\mu (n)+\beta (n)m\right] ,
\end{array}
\label{Ham}
\end{equation}
where $\alpha ^\mu (n)$ and $\beta (n)$ are the four-vector and scalar
matrixes associated with the Dirac matrixes $\alpha ^i$ and $\beta $ in the
privileged frame, namely 
\begin{equation}
\begin{array}{c}
\alpha ^\mu (n)=i\sigma ^{\mu \nu }n_\nu , \\ 
\beta (n)=n_\mu \gamma ^\mu .
\end{array}
\label{marix}
\end{equation}
[This can be easily checked remembering that $(n_\tau )^\mu =(1,0,0,0)$].

\ The parameter $\tau $ is, in general, unrelated to the proper time but, as
we have discussed above, classically one can always choose the coordinate
system in such a way that $\epsilon n_\mu $ can be identified with the
velocity of the particle $\frac{dx^\mu }{ds}=\frac{\pi ^\mu }m$ ($\pi ^\mu
=p^\mu -eA^\mu ).$ Then $\epsilon \tau $ coincides with $s.$ The same
identification cannot be made at the quantum level in general, because the
concept of trajectory is lost. However, in the free case we can argue that
this identification makes sense for the eingenstates of the momentum operator

\begin{equation}
p^\mu \psi _k(x)=k^\mu \psi _k(x).  \label{pk}
\end{equation}
At least in this case, the second member of the second equality in equation (%
\ref{ns}) is well defined.\footnote{%
However, note that this is not the case in the presence of interactions
because we have not any chance to have a common basis which diagonalizes the
four-vector operator $\pi ^\mu $ since $\left[ \pi _\mu ,\pi _\nu \right]
=-ieF_{\mu \nu }.$} By choosing $n_\mu =\epsilon k_\mu /m,$ the first term
in the second member of equation (\ref{Cza}) vanishes, and we finally obtain

\begin{equation}
i\frac{\partial \psi _k(x)}{\partial s}=p_\mu \gamma ^\mu \psi _k(x),%
\hspace{0.5in}s=\epsilon \tau .  \label{Feyk}
\end{equation}
Equation (\ref{Feyk}) resembles the Feynman parametrization of the Dirac
equation (Feynman 1951). However note that the whole formalism discussed
here is restricted to the mass shell ($k^\mu k_\mu =m^2$), due to $\psi
_k(x) $ satisfies the Dirac equation (\ref{D})$.$ In the next section we
briefly discuss the formalism associated with an equation like (\ref{Feyk})
out of the mass shell.

\section{The proper time parametrization of the Dirac equation}

Feynman in 1948, in his dissertation at the Pocono conference (Feynman,
1951; Schweber, 1986; Mehra, 1994), introduced a fifth parameter in the
Dirac equation\footnote{%
The difference between the sign of equation (\ref{Feyk}) and the sign of
equation (\ref{Fey}) is a consequence of having considered different
starting points. While equation (\ref{Feyk}) is a direct covariant
generalization of equation (\ref{HD}) in the hyperplane formalism, equation (%
\ref{Fey}) is motivated by an off-shell proper time formalism, which for the
spatial components preserves the standard results (Aparicio {\it et al}.,
1995a).} 
\begin{eqnarray}
-i\frac{\partial \Psi (x,s)}{\partial s} &=&{\cal H}\Psi (x,s),  \nonumber \\
&&  \label{Fey} \\
{\cal H} &=&p_\mu \gamma ^\mu ,  \nonumber
\end{eqnarray}
to formulate a manifestly covariant (multiple-time) formalism of quantum
electrodynamics.

Equation (\ref{Fey}) is a Schr\"odinger-like equation in which the scalar
Hamiltonian ${\cal H}$ plays the role of a mass operator. Notice that we
retrieve the Dirac equation as an eigenvalue equation, ${\cal H}\Psi
_m=m\Psi _m,$ for stationary states $\Psi _m(x,s)=\Psi _m(x,0)e^{ims}.$ The
evolution operator 
\begin{equation}
U=e^{ip_\mu \gamma ^\mu s},  \label{U}
\end{equation}
is unitary in the indefinite scalar product 
\begin{equation}
\left\langle \Psi ,\Phi \right\rangle =\int \overline{\Psi }(x)\Phi (x)d^4x.
\label{PE}
\end{equation}

The ``norm'' is positive and negative for particle and antiparticle states
respectively (Gaioli and Garcia Alvarez, 1993). Moreover, such
indefiniteness has its root in the St\"uckelberg picture, i.e. it can be
shown that at the semiclassical level (Gaioli and Garcia Alvarez, 1996) 
\begin{equation}
{\rm sign}\left[ \overline{\Psi }(x,s)\Psi (x,s)\right] ={\rm sign}\frac{dx^0%
}{ds}.
\end{equation}

The evolution of any operator $A$ in the Heisenberg picture is given by

\begin{equation}
\frac{dA}{ds}=-i[{\cal H},A],  \label{dbe}
\end{equation}
which is the proper time derivative originally proposed by Beck (1942).

During the last fifty years, this kind of parametrization and the proper
time derivative have been rediscovered or discussed by many authors for
different motivations (Nambu, 1950; Enatsu, 1954; Davidon, 1955; Proca,
1954, 1955; G\"ursey, 1957; Peres and Rosen, 1960; Szamosi, 1961, 1963;
Rafanelli, 1967a, b, 1968, 1970; DeVos and Hilgevoord, 1967; Bunge and
K\'alnay, 1969; K\'alnay and MacCotrina, 1969; Johnson, 1971; Johnson and
Chang, 1974; L\'opez and P\'erez, 1981; Herdergen, 1982; Kubo, 1985; Sherry,
1989; Hannibal, 1991a, 1994; Grossmann and Peres, 1963; Schwinger, 1975;
Rumpf, 1979; Barut, 1988; Barut and Thacker, 1985; Barut and Pavsik, 1987;
Evans, 1990; Fanchi, 1993a, b; Czachor and Kuna, 1997).

In a previous work (Aparicio {\it et. al., }1995a) we have established the
connection between the derivative (\ref{dbe}) and other proper time
derivatives discussed in the literature.\footnote{%
See Fanchi (1993b) for a review of different proposals.} We have concluded
that this is the most satisfactory approach for incorporating the notion of
proper time into the Dirac theory at the quantum level. In other works we
have discussed the interpretation of the formalism (Gaioli and Garcia
Alvarez, 1995, 1996) and the de Sitter invariance of equation (\ref{Fey})
(Garcia Alvarez and Gaioli, 1997a, b). For the sake of completeness, we
review in this Section some points necessary to understand the material
discussed in the previous ones.

We begin by noticing that in equation (\ref{Fey}) the coordinate time $x^0$
has been elevated to the status of an operator, canonically conjugated to
the energy $p^0$. Their commutation relation and the standard canonical one
for the three-position and momentum can be summarized in the covariant
commutation relation

\begin{equation}
\lbrack x^\mu ,p^\nu ]=-i\eta ^{\mu \nu },  \label{xpop}
\end{equation}
which is the quantum analogue of equation (\ref{xp}). It is possible
because, as in the formalism of Section 3, the mass-shell constraint (\ref
{pp}), satisfied by the irreducible representations of the Poincar\'e group,
is no longer valid. In this case, there is a new dynamical group of
symmetries that enlarges the Poincar\'e group, that is, the de Sitter group,
which could have been taken as the starting point to obtain the Feynman
parametrization (Garcia Alvarez and Gaioli, 1997a, b). Here we have followed
the heuristic argument given in Section 4 to obtain the form of a covariant
Hamiltonian conjugated to the proper time $s$ on the mass shell and after
that, we have extrapolated this form out of the mass shell. We conclude this
Section giving an independent argument which shows that the operator $p_\mu
\gamma ^\mu $ determines the evolution of the system in the proper time $s$.

Using (\ref{dbe}), for $A=x^\mu ,$ and (\ref{xpop}) we obtain the covariant
generalization of Breit's (1928, 1931) formula

\begin{equation}
\frac{dx^\mu }{ds}=\gamma ^\mu .  \label{Breit}
\end{equation}

Projecting this equation of motion on positive and negative mass states, for
eliminating the covariant {\it Zitterbewegung,} by means of the projectors

\begin{eqnarray}
\Lambda _{\pm } &\equiv &\frac 12\left( 1\pm \frac{{\cal H}}{\sqrt{{\cal H}^2%
}}\right) ,  \label{p} \\
&&  \nonumber \\
\Lambda _{\pm }{\cal H}\Lambda _{\pm } &=&\pm \Lambda _{\pm }\sqrt{p^\mu
p_\mu }\Lambda _{\pm },  \label{pHp}
\end{eqnarray}
we have

\begin{equation}
\Lambda _{\pm }\frac{dx^\mu }{ds}\Lambda _{\pm }=\pm \Lambda _{\pm }\frac{%
p^\mu }{\sqrt{p^\mu p_\mu }}\Lambda _{\pm }.  \label{pdxp}
\end{equation}
The projected Hamiltonian and four-velocity are analogues of (\ref{J}) and (%
\ref{dxJ}) respectively, which on the positive mass shell leads us to the
identification of the evolution parameter with the proper time. Moreover, we
see that eliminating the interference between positive and negative states
we have the analogue of the proper time constraint (\ref{constr}), namely

\begin{equation}
\Lambda _{\pm }\frac{dx^\mu }{ds}\Lambda _{\pm }\Lambda _{\pm }\frac{dx_\mu 
}{ds}\Lambda _{\pm }=\Lambda _{\pm }.  \label{pdxp1}
\end{equation}

\section{Further remarks and conclusions}

Summarizing, the standard canonical formalism has two difficulties:

$a$) It does not provide a relativistic invariant description of the
dynamical evolution of the system,

$b$) It does not enable us to include simultaneously particles and
antiparticles states.

The problem ($a$) is because the coordinate time is not a Lorentz scalar,
and {\it (b)} is due to the fact that particles and antiparticles go
forwards and backwards in this time respectively. Then the coordinate time
is not able to describe processes involving both species simultaneously. The
standard canonical formalism of quantum field theory is a many particle
formalism with negative and positive charges for the particles and
antiparticles respectively.\footnote{%
This double sign of the charge has its correlate in the double sign of the
kinetic energy, $\left( \frac{dx^0}{ds}=\epsilon \frac{\sqrt{m^2+{\bf p}^2}}%
m\right) ,$ in the Feynman-St\"uckelberg picture, while the sign of the
energy and the charge is kept unaltered in the standard picture and in the
St\"uckelberg one, respectively.} Such a picture reinterprets the notion of
antiparticle of the St\"uckelberg picture by reversing the direction of its
space-time trajectory, which is equivalent to conjugate its charge.\footnote{%
This property, emphasized by Feynman (1948, 1949) at the classical level, is
also held in the quantum case (Garcia Alvarez and Gaioli, 1997b).}

The first difficulty ($a$) is removed by the Fleming formalism, but there is
a price to be paid.

$a^{\prime })$ It has an arbitrariness in the choice of the privileged
system.

We have shown that as soon as we try to remove this arbitrariness, by
choosing $n^\mu =\epsilon \frac{dx^\mu }{ds},$ we get to the proper time
formalism on the mass shell. But in this case difficulty ($b$) still
remains. We have to label the dynamics with the time $s=\epsilon \tau $ $($%
in this case $\tau $ is the proper time of the particle) for having the same
label for both particles and antiparticles: A solution which naturally
arises in the proper time formalism out of the mass shell.

The last discussion suggests us how to remove difficulty ($b$) at the level
of the hyperplane formalism. One could label the dynamics with $\xi =$ $%
\epsilon \tau .$ In this case the Hamiltonian, corresponding to equation (%
\ref{Hn}) and to the Dirac equation in the hyperplane, reads

\begin{equation}
H_\xi (n)=\epsilon n^\mu p_\mu ,  \label{Hnep}
\end{equation}

\begin{equation}
i\frac{\partial \psi (x)}{\partial \xi }=\epsilon n_\mu (i\sigma ^{\mu \nu
}p_\nu +m\gamma ^\mu )\psi (x).  \label{Czaep}
\end{equation}
Note that the new Hamiltonian, as the rest mass, becomes definite positive.
The hyperplane formalism corresponding to equation (\ref{Hnep}) out the mass
shell is equivalent to the Hall and Anderson formalism, identifying $t^\mu $
with $\epsilon n^\mu .$ Moreover, it is interesting to see the analogy
between (\ref{Hnep}) with the covariant Hamiltonian (\ref{Fey}), identifying 
$\epsilon n^\mu $ (which temporal component is $n_\tau ^0=\epsilon )$ with $%
\gamma ^\mu $ (notice that the eingenvalues of $\gamma ^0$ are $\pm 1).$

Finally, in contrast with the standard case, the scalar product associated
with the new Dirac equation (\ref{Czaep}) is indefinite, i.e.

\begin{equation}
\left\langle \psi ,\psi \right\rangle =\int \overline{\psi }\gamma _\mu \psi
d\sigma _\xi ^\mu =\epsilon \int \overline{\psi }\gamma _\mu \psi n^\mu
d\sigma _\tau .  \label{PEep}
\end{equation}
As in the case of equation (\ref{PE}), this indefiniteness arises as a
consequence of describing particle and antiparticle dynamics with the same
label.

To summarize, the relation between the standard canonical picture and the
Feynman-St\"uckelberg one can be synthesized as follows:

In the first one the mass, the kinetic energy, and the scalar product are
always positive definite. Both particles and antiparticles go forward in
coordinate time and proper time, and they are only distinguished by the sign
of the charge. In the second case, both particles and antiparticles have
positive mass, but only the proper time evolution goes forward for both
species. Particles and antiparticles have positive and negative kinetic
energy respectively, and according to this they go forwards and backwards in
coordinate time. The charges are the same for both species, which avoids the
use of a many particle formalism in order to describe particle creation and
annihilation processes.\footnote{%
Remember that, at the semiclassical level, such processes can be pictured as
a zig-zag trajectory in space-time (Feynman, 1948, 1949; Garcia Alvarez and
Gaioli, 1997).} As a consequence we also have and indefinite scalar product,
something which strikes against our standard notions. Moreover it has been
the historical reason for which Dirac disregarded the Klein-Gordon equation
(Weinberg, 1995). But, like the double sign in the energy, it has its roots
in the indefinite metric of the Minkowski space-time manifold (\ref{ds2}).%
\footnote{%
Notice that an indefinite metric space also appears in the covariant
quantization of electromagnetic fields (Gupta, 1950; Bleuler, 1950; see also
Jauch and Rohrlich, 1976, Chap. 6 and Cohen Tannoudji {\it et. al.,} 1989),
Chap 5).} In other words, while the second picture seems to be the natural
way for adapting the principles of quantum mechanics to the theory of
relativity, the first one looks as a deliberated attempt for keeping our old
picture of non-relativistic quantum mechanics for describing the full
relativistic quantum phenomena.

\end{document}